\documentclass[aps,pre,twocolumn,groupedaddress,showpacs,amsfonts,10pt,%
tightenlines,floatfix]{revtex4-1}

\usepackage[utf8]{inputenc}
\usepackage{graphicx}
\usepackage{amsmath,amssymb}
\usepackage{upgreek}
\usepackage[usenames,dvipsnames]{xcolor}
\usepackage{braket}
\usepackage{cancel}

\graphicspath{{./Abbildungen/}}

\begin{document}

\title{Universal computation using localized limit-cycle attractors in neural networks}

\author{Lorenz Baumgarten}
\email[]{lbaumgarten@itp.uni-bremen.de}

\author{Stefan Bornholdt}
\email[]{bornholdt@itp.uni-bremen.de}

\affiliation{Institut für Theoretische Physik, Universität Bremen, 28359 Bremen, Germany}

\date{\today}

\begin{abstract}
Neural networks are dynamical systems that compute with their dynamics. 
One example is the Hopfield model, forming an associative memory 
which stores patterns as global attractors of the network dynamics. 
From studies of dynamical networks it is well known that localized 
attractors also exist. Yet, they have not been used in computing paradigms. 

Here we show that interacting localized attractors in threshold networks 
can result in universal computation. We develop a rewiring algorithm that 
builds universal Boolean gates in a biologically inspired two-dimensional 
threshold network with randomly placed and connected nodes using collision-based computing.
We aim at demonstrating the computational capabilities and the ability 
to control local limit cycle attractors in such networks by creating 
simple Boolean gates by means of these local activations.
The gates use glider guns, i.e., localized activity that periodically 
generates "gliders" of activity that propagate through space.
Several such gliders are made to collide, and the result of their 
interaction is used as the output of a Boolean gate.
We show that these gates can be used to build a universal computer.
\end{abstract}


\maketitle

\section{Introduction}
Computation in nature occurs in highly irregular environments that differ significantly 
from regular human-constructed computation methods. In this spirit, 
the field of unconventional computing \cite{Toffoli1998, Advances1, Advances2}  
explores alternative methods of computation to the ubiquitous von-Neumann architecture 
of modern computers. A common and promising strategy is using biology as inspiration 
for new computation schemes, as in the field of neuromorphic computing \cite{Schuman2017}, 
and the sub-field of amorphous computing with its large numbers of irregularly spatially 
distributed, unreliable, and locally communicating parts \cite{Abelson1995, Nagpal2004, Abelson2009}.
Such irregularly placed and only partially connected parts can, for example, be found in neural brain networks. 
\\
Unconventional computing schemes find uses in a variety of fields such as managing robot swarms 
\cite{Otte2018, Hamann2016}, engineering biological devices \cite{Macia2016}, medical image analysis 
\cite{Mitra2015}, or information storage \cite{Nugent2008}, and a multitude of other ideas, 
such as cellular neural networks \cite{Chua1988}, for example, have been developed.

In particular, highly parallelizable computation networks, such as memristor networks 
\cite{kozma2012, adamatzky2013, vourkas2016, chua2019}, that can be trained like artificial neural networks 
\cite{nugent2014, yakopcic2017}, appear highly promising. We take recent advances in this field as inspiration 
for creating a new unconventional computing scheme in irregular, randomly constructed neural networks.

A major mechanism of computation in neural networks is computing with attractors, where the global attractors 
of the dynamical network represent the result of a computation \cite{Murali2018, Aravindh2018}. 
This computing paradigm is perhaps best exemplified by the Hopfield model \cite{hopfield1982} 
in which patterns are stored as global attractors of the network dynamics.
It has long been discussed that computation in the brain takes advantage of using 
attractors, including non-fixed point (or limit-cycle) attractors \cite{Hertz1995}. 

One prominent property of attractors in asymmetric neural networks is that, 
under certain circurmstances, they may occur as localized excitations. 
 Such localized attractors, or localized persistent activity, have been observed in neural networks 
 \cite{samsonovich1997, ermentrout1998, hansel1998, sharp2001, wang2001, brunel2003, roudi2004, 
 rubin2004, schrobsdorff2005,  koroutchev2006, gonzalez2011, Monasson2014}, and have been 
 discussed in diverse systems, such as genetic networks \cite{kauffman1984, kauffman1993, kauffman2003}
 and immune networks \cite{weisbuch1990, neumann1992, weisbuch1994}, 

We here expand the idea of attractor computation to co-existing, localized attractors. 
In an example system, we use multiple spatially localized periodic (or limit-cycle) attractors, 
as opposed to the conventionally used global attractors in artificial neural networks such as the Hopfield model. 

As a proof of concept, to demonstrate the possibility of localized attractor computation in irregular 
neural networks, we will make use of collision-based computing, which utilizes moving particle-like 
localized activity islands, as have been observed in attractor neural networks in \cite{Monasson2014}.
We do not, however, suggest that the algorithm and resulting dynamics described in this paper accurately 
reflect a brain's function; we merely propose a biologically inspired new unconventional computing method.
\\
Collision-based computing is the computation of signals propagating through space, usually called gliders, 
solitons, or wave-fragments depending on context, by interaction on impact with each other or obstacles.
It is the subject of research in a variety of different systems such as non-linear 
\cite{Jakubowski1996, Jakubowski2017, Martinez2012} and chemical media such as the Belousov-Zhabotinsky medium 
\cite{Toth1995, Toth2009, Adamatzky2004, Steinbock1996, Costello2009} and liquid marbles \cite{Draper2017}, 
and biological systems such as biopolymers \cite{Siccardi2016, Siccardi2017, De2016} and slime molds 
\cite{Adamatzky2010, Jones_Adamatzky2010, Adamatzky2011}, see 
\cite{Advances1, Advances2, CollComp, Adamatzky2012, NonlinearMediaComp} for reviews.
\\
The field emerged in the wake of Fredkin and Toffoli's paper \cite{Fredkin_Toffoli1982} introducing 
the idea of a ballistic computer---the billiard ball model---, in which Boolean logic gates were 
implemented by collisions between billiard balls and reflectors; Margolus' following paper \cite{Margolus1984} 
creating a cellular automaton implementation of the billiard ball model; and Berlekamp, Conway, and Guy 
creating Boolean logic gates using gliders in the game of life \cite{WinningWays2}.
Since then, various other collision-based computing schemes for cellular automata
\cite{Squier1993, Jakubowski1996, Zhang2009, Sapin2007, Hordijk1998, Jakubowski2017, Adamatzky1998, 
Adamatzky2006, Martinez2012, Martinez2018} or in preconstructed mazes \cite{Becker2019} have been developed .
Unlike our systems, however, these automata operate on regular lattices.
\\
We demonstrate how limit cycles can be manipulated by rewiring algorithms to achieve desired results.
For this, we will create Boolean gates operating on limit cycle glider guns and show that universal computation 
using these gates is possible.

\section{Model}
We study a network of $N$ nodes randomly distributed in a two-dimensional square of space whose side length we define as 1.
The nodes have directed connections between each other in such a way that the probability $P$ of a connection 
existing from node A to node B is proportional to an exponential function
$$
P(d) = K\exp(-\lambda d)
$$
of the distance $d$ between A and B.
The parameters $K$ and $\lambda$ are chosen to result in specific values for the average degree $k$ 
and the clustering coefficient $C$. We choose a relatively high clustering coefficient and average degree due to our observations of localized attractors in the networks we studied in \cite{baumgarten2019}.
\\
Nodes are either excitatory or inhibitory, meaning that, if they are active, they send a positive or negative signal 
to all nodes they have efferent connections to. A node $i$'s state $\sigma_i$ is determined by its incoming signal
$$
S_i = \sum_j c_{ij} \sigma_j(t)
$$
via
$$
\sigma_i(t+1) = \begin{cases} 1 & \text{if~} S_i>h\\
	0 & \text{otherwise}
\end{cases},
$$
where $c_{ij}$ is $\pm1$ if there is a connection from node $j$ to node $i$ and zero otherwise, and $h$ is the threshold.
All nodes are updated synchronously in discrete time steps.

In all our simulations, we use an initial network with $N=2000$ nodes, threshold $h=2$, average degree $k=10$, clustering coefficient $C\approx0.4$, and a chance of nodes being excitatory or inhibitory of 50\,\% each. 
In Movie S1 in the supplemental material, we show an animation of localized attractors in a similar, untrained random network.
\\
To create logic gates, we will encode incoming signals in glider guns that periodically produce propagating 
patterns called gliders. These gliders will collide and interact with each other to produce a desired output.
\\
Let us first discuss how glider guns are created and afterwards discuss two different strategies to utilize 
these glider guns for logic gates.

\section{Glider guns}
\begin{figure}
	\includegraphics[width=0.8\linewidth]{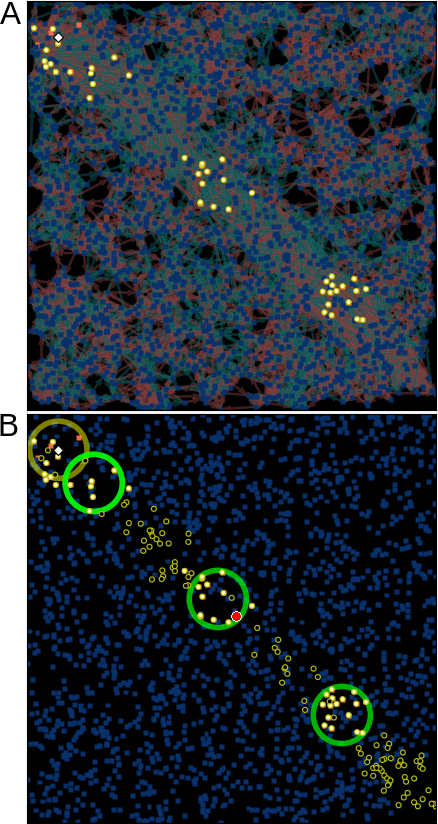}
	\caption{A glider gun with period $7$ created by the algorithm discussed above. 
	Blue and red squares are nodes that are either off or on throughout the entire limit cycle, 
	respectively. Yellow circles indicate nodes that change their state at least once during 
	the limit cycle and are filled if the node is currently on; currently inactive nodes 
	that change their state are omitted in (A). White diamonds indicate active input nodes. 
	In (A), the underlying connections are shown: Red and blue arrows indicate excitatory 
	and inhibitory connections, respectively. 
	In (B), region II is marked by a yellow ring around the input, 
	region III is marked by green rings, and the target point is indicated by a red dot.}
	\label{GliderGun}
\end{figure}

To create a glider gun, we denote a node as an input node whose state will be defined from outside 
instead of by the network dynamics and which will serve to activate a glider gun, meaning it will 
periodically produce an activation that will propagate through space. This node will send a signal 
of strength $h+1$ instead of strength one to the nodes it is connected to, i.e., $c_{ij}=h+1$ or $c_{ij}=0$ 
where node $j$ is the input node, so that its signal is sufficient to activate nodes in its vicinity 
given no other incoming signals.
\\
We also define a target point towards which the glider will move with constant velocity within $T$ time steps.
The glider need not necessarily stop at the target point; therefore, the target point only defines 
a glider's direction and speed, not its destination. Throughout this paper $T$ will be chosen as $T=10$.
\\
To set a glider gun's corresponding period, we rewire connections in an area around the input node randomly 
and measure the period of the limit cycle that is reached when initially only the input node is active. 
If the resulting period is further from the desired result than before rewiring, the rewiring is undone. 
This is repeated until the desired period is produced.
Here, and throughout this paper, rewiring is done by choosing two connections and swapping their target nodes 
\cite{MaslovSneppen}, 
so long as that does not result in redundant connections between two nodes, preserving all nodes' degrees.
Rewirings are also only done if they do not result in connections above a certain length $L$ 
to preserve the network's spatial character.
\\
To now create gliders, we divide space into three regions: region I in which we do not want activity, 
region II in which we do want activity and region III where anything is allowed to happen. 
We define a fitness function $f$ as
\begin{align*}
	f &= \frac {\sum_\mathcal{A}\sum_i g(S_i, x_i, y_i)}{|\mathcal A|} ~\text{with}\\
	g(S, x, y) &= \begin{cases}
		\min( h-S, 0 ) &\text{if $(x,y)$ in I}\\
		\max( S-(h+1), 0 ) &\text{if $(x,y)$ in II}\\
		0 & \text{otherwise}
	\end{cases}.
\end{align*}
where $(x_i, y_i)$ are the node $i$'s coordinates, $\mathcal A$ is the set of network states in the 
network's limit cycle, and $|\mathcal A|$ is the number of states in the limit cycle.
\\
Now the three regions need to be defined:
Region III is the region immediately around the input node, with a radius $D$, for which we choose 
$D=0.07$ in our simulations. The length $D$ also governs the maximum length of formed connections $L=3D$.
For the glider gun to periodically produce gliders, some periodic activity is required, and it 
does not make sense to promote or suppress activity here.
\\
Region II are the gliders themselves. It consists of areas of radius $D$ that are periodically created 
at the input node and move with fixed velocity towards and past the target point. 
The initial position of these areas is chosen to maximize the fitness produced by region II.
If, for example at the start of a new glider shot, region II and III overlap, 
nodes in the overlap are counted as in region II.
\\
Region I is the rest of space.
\\
Now, we start with a completely deactivated network and activate the input node.
Then, the network dynamics are run until a limit cycle is reached, and the fitness within the limit cycle is calculated.
Afterwards a rewiring operation is done, and the previous calculation is repeated with the same starting conditions.
If the limit cycle's period changes, no limit cycle is found within a set amount of time steps, 
the network does not return to an inactive state after deactivating all input nodes at a random point in the limit cycle, 
or the fitness is lower after rewiring, the rewiring is undone.
This is repeated until a satisfactory glider gun has been created. 
Because we want our computations to function regardless of when input nodes are activated or deactivated, 
between two of such rewiring attempts, we deactivate the input node and wait a random number of time steps 
before reactivating the input node and measuring fitnesses. Here, and in the rest of this paper, 
a random number of time steps is always a number between zero and the end of the first limit cycle that is reached.
\\
An example of such a glider gun is shown in Figure \ref{GliderGun}.

\section{Logic Gates}
In this section, we will discuss how to utilize glider guns to build logic gates.
For this, we have developed two different strategies.
\\
Both strategies use multiple input nodes activating glider cannons that aim at the same target point. 
The gliders will therefore collide at the target point and interact with each other. 
This interaction will define the output.
\\
We want our gates to function regardless of when the input nodes are activated, 
and therefore we use glider cannons with different prime number periods. 
The reasoning behind this is that, using two primes, all possible phase differences between a state 
of one glider gun and a state of another glider gun will occur at some point.
\\
This is not exactly correct since, once the glider guns interact with each other, their periods 
will not be as clear cut as before. Instead, a macro limit cycle involving the dynamics of 
all active glider guns will be created. In an effort to force the individual glider guns to retain 
their periodic behavior, we will only accept rewiring if, when multiple glider guns are active, 
the macro limit cycle's period is the same period as if the individual glider guns were not 
interacting with each other. This macro period is the least common multiple of individual periods, 
which is, since we are using prime numbers, the product of the periods.
\\
In the first strategy, we define one output for every input node.
An output is counted as TRUE if the glider passes the target point and FALSE otherwise.
In the second strategy, we define a common output region with width $D$ for multiple glider cannons.
Since for a common output the glider gun signals have to merge at the target point and since 
the glider guns have different periods, it is not clear which period the output signal should have.
Therefore, instead of a signal moving towards the output region, we will assign the area 
between the target point and output region statically to region II if a positive output is required.
\\
For both these strategies, the area of radius $D$ around the target point is also added to region III, 
meaning any behavior is permitted here. This allows, for example, the signal from one glider gun 
to remain within this region to then catch a signal from another glider gun and interact with it 
without the need for the signals to arrive simultaneously.
\\
Also, for strategy two and for gliders in strategy one whose output is either supposed to be FALSE 
or for whom no desired output is defined, the glider shots are terminated at the target region, 
while these shots' region II is overwritten by the target region's region III.
This means that these shots are only forced into existence outside the target region.
This, for example, makes it easier for an AND-interaction to occur because otherwise 
both signals would compete for activating all nodes in the interaction region by themselves, 
as opposed to only in the case when both signals are present.
\\
Both these strategies have advantages and disadvantages:
\\
For strategy one, if the desired output requires only one input node to be active, 
say an $A\neg B$ gate, the resulting output has the period of the active glider gun 
and can therefore simply be routed towards another gate for further computations. 
Since the $A\neg B$ gate is universal, any Boolean operation can be created using this principle. 
On the other hand, if an output requires multiple symbols to be active, 
the resulting output signal will in general have the rather large and unwieldy period 
of the macro limit cycle. This output will likely need to be read out and converted 
into a new input signal to start a new glider gun. This could be accomplished by simply 
setting nodes at the glider's end to permanently be in region II and therefore be able 
to permanently activate a glider gun---fig. \ref{Snapshots} (D) shows that permanent 
activity of nodes in such an area is possible---, given that the previous input remains, 
or by defining an input node that only turns off when it has not received a signal 
for a period of time longer than the macro limit cycle.
\\
For strategy two, the same period length problems apply. For this strategy's advantages, 
let us discuss how one would create an XOR-gate using the two separate strategies:
\\
For strategy two, the common output region can simply be trained to output XOR, 
removing the need to create more complicated circuits.
For strategy one, an XOR-gate cannot be realized on one of the outputs since the output 
belonging to an input node A can only be TRUE if there is an incoming signal from input node A. 
Therefore, one has to either reroute the outputs of an $A\neg B$- and a $B\neg A$-function 
to the same output region and add them together or  use the possible gates, for example 
the universal $A\neg B$ gate, to build a circuit with an XOR output.
\\
Fortunately, multiple gates can easily be combined.  The direction and speed of different 
gliders in our algorithm is simply constant for convenience's sake; however, nothing dictates 
that a glider cannot change direction or speed, and therefore it is easily possible to reroute 
signals to arbitrary points in the network or to delay or accelerate them, should it be required.

\section{Algorithm}
The algorithm to create gates is similar to the one for creating glider guns, but needs 
to be expanded to deal with various issues that can occur when multiple input nodes are active.
\\
Note that we will distinguish between inputs and input nodes. 
An input is one combination of active or inactive input nodes.
\\
Firstly, we need to ensure that the gate works correctly for all possible inputs, so the calculation 
of the network's fitness will now consist of activating some combination of input nodes, 
measuring the fitness for this input, and repeating this process for all possible inputs 
(excluding all input nodes being inactive).
Again, between different inputs, all input nodes are deactivated and the network dynamics 
are run for a random number of time steps before the next input is activated.
The final fitness is then the sum of fitnesses for the individual inputs.
\\
One important property we want our gates to have is for them to function regardless of when and 
in which order input nodes are activated.
Therefore, instead of simultaneously activating all input nodes in a specific input, 
individual input nodes are activated in random order and with a random number of time steps between them.
With the random number of time steps any possible glider gun interaction, during the previously active 
guns' transient dynamics or within the limit cycle, can occur.
\\
Because of this large number of possible activation patterns, it is unreasonable to calculate 
the fitness for all of them for every rewiring attempt; instead, we only calculate whether 
the fitness increases for one set of activation patterns per rewiring attempt.
This, unfortunately, may lead to rewirings worsening the fitness for different activation patterns.
This can also lead to the macro limit cycle's period changing or the network not returning 
to an inactive state without inputs.
When it is detected that either of those two happened, previous rewirings are sequentially undone 
in reverse order until the problem no longer occurs for the activation pattern for which this was detected.
\\
Another issue that may occur is that some activation patterns may have a significantly 
lower fitness than others, and a rewiring that improves this pattern will often lower other 
activation patterns' fitnesses to a similar value. To avoid this, an activation pattern 
that has a significantly lower fitness than the previous pattern will be skipped.
\\
To speed up the rewiring, when searching for a valid rewiring step, 
activation patterns are reused until a rewiring step that actually improved --- instead of 
just preserving --- the fitness is found, so as to not be forced to recalculate 
the fitness before rewiring at every step.
\\
When choosing connections to rewire, one of the connections chosen has to originate from a node 
that at any point during the calculation of the fitness, during the transient or the limit cycle, 
has been active to further speed up the algorithm, since rewiring connections that do not 
transmit any signal has no effect.
\\
Additionally, not all connections in the network are considered for rewiring.
Instead, the lowest distance in the direction to the target point that any cannon shot 
has reached during any of the inputs normalized by the distance between the corresponding 
input node and the target point is calculated. Only connections which lead to nodes within 
a region depending on this distance are considered for rewiring.
The algorithm alternates between choosing this region as a region around the points that lie 
at this minimum distance in the direction from the input nodes to the target point with 
radius $D$ and as the entire path of the gliders up to those points.
When calculating this distance, cannon shots that are not meant to pass the target point 
are disregarded as long as they get close enough to the target point.
For shared outputs, once all cannon shots get close enough to the target point, 
the minimum reached distance from the target point to the end of the output region 
for an input with desired output TRUE is used instead.
Alternating between these two regions has the advantage that, for the region around the minimum 
distance point, there is a good chance for the rewiring to result in the cannon shot traveling 
farther after rewiring while the other region can optimize the path that has already been created.
\\
Also, and this is vital for the algorithm to function, by only rewiring up to the lowest 
distance reached, when multiple cannon shots have to pass the target point, 
a situation in which one cannon shot already reaches far past the target point while 
the other has not passed the target point yet is not created. In such a situation, 
any rewiring around the target point necessary to make the second shot pass the target point, 
that would negatively affect the first shot would significantly lower the fitness because 
it would cut off the first shot significantly earlier than before while only slightly 
increasing the distance that the second shot travels. In such a situation, it is difficult 
to find a rewiring that improves the second shot without ruining the already established first shot.
\\
Finally, when activation patterns are skipped because they have a significantly lower fitness 
than previous patterns, after skipping configurations 100 times in a row, it is assumed 
that something has gone wrong and previous rewirings are sequentially undone similar to when 
an activation pattern results in the wrong period, until the problem does not occur any longer.

\section{Results}
\begin{figure*}
	\includegraphics[width=0.87\linewidth]{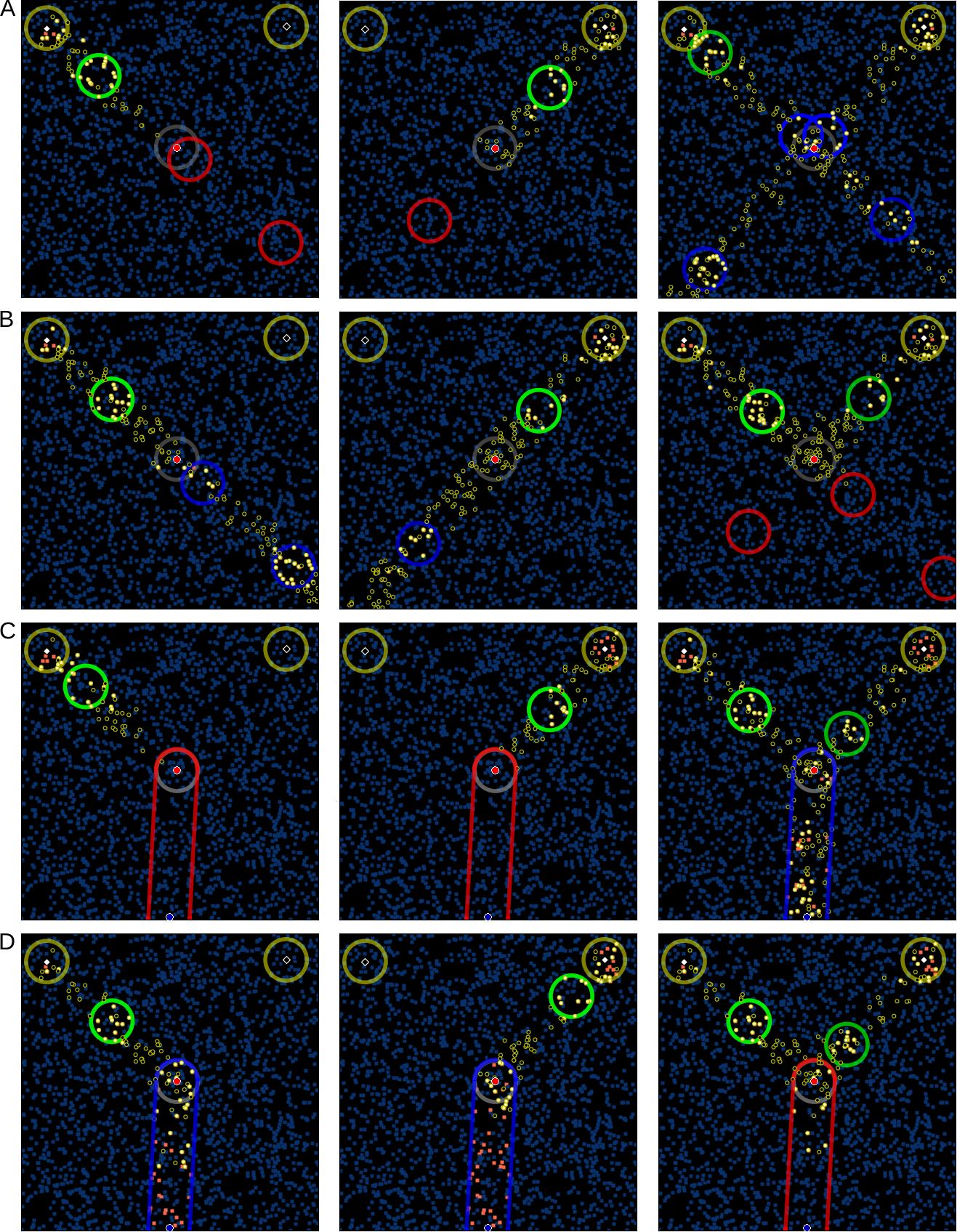}
	\caption{ Snapshots of (A) an AND-gate for both input nodes, 
	(B) a gate with an $A\neg B$ output for input node $A$ and a $B\neg A$ output for input node $B$, 
	(C) a common AND output, and 
	(D) a common XOR output. 
	In all of these, the left glider gun has period seven, and the right one has period eleven. 
	Black diamonds denote deactivated input nodes. Gray circles indicate the region III around the target point. 
	In (A) and (B), red rings merely indicate where a shot had been, 
	had it not been stopped at the target point, and belong to region I, 
	and blue rings indicate shots that were supposed to pass the target region and belong to region II. 
	In (C) and (D), red and blue lines indicate the areas that are fixed as regions I or II, 
	respectively, and blue circles mark the center of the output region. 
	Remember that gray rings overwrite green and red areas and are in turn overwritten by blue areas. 
	The movies S2--S13 in the supplemental material show animations of the gates shown.}
	\label{Snapshots}
\end{figure*}

In this section, we will present the results of an AND-gate on both outputs and an $A\neg B$-gate 
on one output and a $B\neg A$-gate on the other one for strategy one as well as an AND- and an 
XOR-gate for strategy two. Snapshots for all these gates are shown in Figure \ref{Snapshots}, 
and the fitness as well as the error rate as a function of rewiring attempts is shown in 
Figure \ref{fitness}. 
Animations of these gates can be found in the supplemental material, movies S2--S13.

\begin{figure}
	\includegraphics[width=1\linewidth]{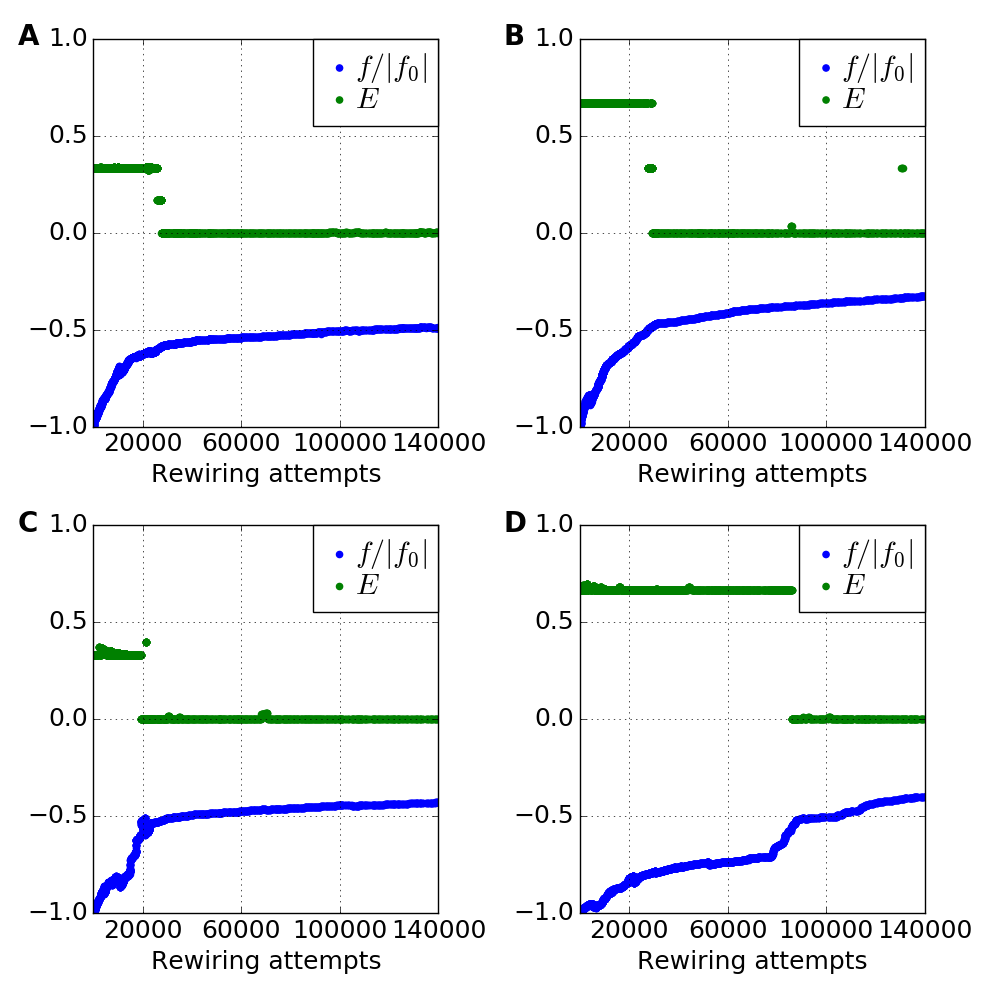}
	\caption{ Fitness $f$, normalized by initial fitness $f_0$, and error rate $E$ as a function of rewiring attempts for 
	(A) an AND-gate for both input nodes, 
	(B) a gate with an $A\neg B$ output for input node $A$ and a $B\neg A$ output for input node $B$, 
	(C) a common AND output, and 
	(D) a common XOR output. 
	The initial error rate for example for (A) is $1/3$ because initially the cannon shots cannot reach the output region, 
	which is the wanted result for inputs in which only one input node is active. 
	Therefore, the result is initially correct for two out of three possible inputs.}
	\label{fitness}
\end{figure}
By periodically measuring error rates under the same conditions used during rewiring, i.e., 
activation of input nodes at random times and in random orders and with random numbers of time steps 
between inputs, and stopping the rewiring algorithm when minimal error rates are achieved, 
it is easily possible for all these gates to achieve perfect performance. The results of an input 
is counted as an error if any of the outputs is not the desired result, if the macro limit cycle's period 
differs from the intended period, or if the network would not return to the deactivated state after 
deactivating all input nodes. When measuring error rates, all possible inputs are used equally frequently, 
except for the zero input, which is already implicitly covered by the third error condition.

\section{Conclusion}
We have demonstrated the possibility of computation with attractors in irregular two-dimensional 
threshold networks. For this, we constructed a rewiring algorithm that enables us to control the 
behavior of localized limit cycle attractors within such networks. With this algorithm, we first 
created glider guns to propagate signals in space and then used these glider guns to build Boolean gates. 
We have developed two strategies for such gates, both involving the collision of multiple gliders 
from different glider guns. 
In the first strategy, every glider gun has its own associated output, whereas in the second strategy, 
the entire gate only has one common output.
\\
We have built multiple Boolean gates with either of these strategies and argued that these gates 
can easily be combined to build a universal computer.
We have also demonstrated that these gates can achieve perfect performance, in the absence of noise, 
even given random activation and deactivation times of the incoming inputs.
\\
This is, to our knowledge, the first application of localized activity in such networks, 
and we hope that it may therefore be useful to gain insight on the operation of brain networks 
in which localized activity as a response to external stimuli can also be observed.
\\
Further, the computation method described in this paper is merely one option for utilizing 
localized attractors for computation in threshold networks. A number of different computation 
schemes are also conceivable and may hopefully be explored in the future.
We hope that this simple demonstration can spark new ideas for amorphous computation schemes 
using localized activity in neural network structures.
\\
The gates shown here would most likely not work if the updates to the nodes were not synchronized
or if the signals or node states were subject to noise, and neither such synchronization 
nor a noise-free environment are to be expected in real-world applications. However, in biology, 
genetic networks can reliably function under such conditions \cite{klemm2005, braunewell2007}, and
another interesting model for reliable behavior from noisy elements has been demonstrated recently 
with the game of life, a cellular automaton with gliders similar to those used in this work, 
which has been successfully implemented to reliably function in a noisy environment \cite{chan2018}. 
Both these findings point towards the possibility of reliably managing network dynamics like 
the ones presented in this paper in the presence of noise.
\\
Besides the question of noisy implementations of our model, a second line of possible future research 
is the evolutionary creation of the network itself. The algorithm presented here is a stepwise evolutionary 
algorithm, using mutation and subsequent selection in an overall algorithmic process. An interesting question 
is how a developmental algorithm, perhaps on the basis of only locally available information, could address the problem. 

\bibliography{references}

\section*{Supplemental Material}
This section contains explanations of the supplementary movie files.
\subsection*{Movie S1}
Local attractors occur in a two dimensional irregular neural network at large clustering coefficient $C$. We show an animation of a random network with $k=40$, $N=4000$, $h=2$, $C=0.5$, and a probability of nodes being excitatory or inhibitory of 50\,\% each. To better illustrate the spatially disjoint nature of the attractors, only connections between nodes are shown whose states change in the cyclical attractor. 

\subsection*{Movies S2--S13}
The movies S2--S13 show animations of the gates shown in Figure \ref{Snapshots}. The possible combinations of active input nodes are shown in a separate movie each, resulting in three movies per gate.
Movies S2--S4 show a gate with an AND output for both input nodes; movies S5--S7 show a gate with an $A\neg B$ output for input node $A$ and a $B\neg A$ output for input node $B$; movies S8--S10 show a gate with a common AND output; movies S11--S13 show a gate with a common XOR output.
\end{document}